\documentclass[prc,twocolumn,showpacs,showkeys,nofootinbib]{revtex4}
\usepackage{graphicx}  
\usepackage{dcolumn}   
\usepackage{bm}        
\usepackage{amsmath, amsthm, amssymb}

\usepackage[bookmarks]{hyperref}

\def\P{\Phi}
\def\AP{\bar{\Phi}}

\begin{document}

\title{Strange matter and kaon to pion ratio in SU(3) PNJL model}

\author{A. V. Friesen}
\affiliation{Joint Institute for Nuclear Research,  Dubna,
Russia}

\author{Yu. L. Kalinovsky}
\affiliation{Joint Institute for Nuclear Research,  Dubna,
Russia}

\author{V. D. Toneev}
\affiliation{Joint Institute for Nuclear Research,  Dubna,
Russia}

\begin{abstract}
The behavior of strange matter in the frame of the SU(3)Polyakov-loop extended Nambu-Jona-Lasinio model including $U_A(1)$ anomaly is considered.  We discuss the appearance of a peak in the ratio of the number of strange mesons to non-strange mesons known as the "horn". The PNJL model gives a schematic description of the chiral phase transition and meson properties at finite temperature and density.  Using the model, we can show that the splitting of kaon and anti-kaon masses appears as a result of introduction of density. This may explain the difference in the $K^+/\pi^+$ ratio and $K^-/\pi^-$ ratio at low $\sqrt{s_{NN}}$ and their tendency to the same value at high $\sqrt{s_{NN}}$. We also show that the rise in the ratio $K^+/\pi^+$ appears near CEP when we build the $K^+/\pi^+$ ratio along the phase transition diagram and it can be considered  as a critical region signal.
\end{abstract}


\maketitle

\section{Introduction}
Now the study of the matter formed during the collision of heavy ions at high energies is on top of interest of high energy physics. Much interest still belongs to the search of the critical end point and phase transition in the hot and dense matter. The search for quark-gluon plasma (QGP)  where hadrons dissolve into interacting gluons and quarks is difficult due to the short lifetime of the QCD phase. It is needed to find sensible probes for the transition to the QGP phase (i.e. deconfinement transition and the chiral symmetry restoration). One of the suggested signals was the strangeness enhancement which was explained through the interactions between partons in QGP.

\begin{figure}[h]
\centerline{
\includegraphics[width = 8.5 cm]{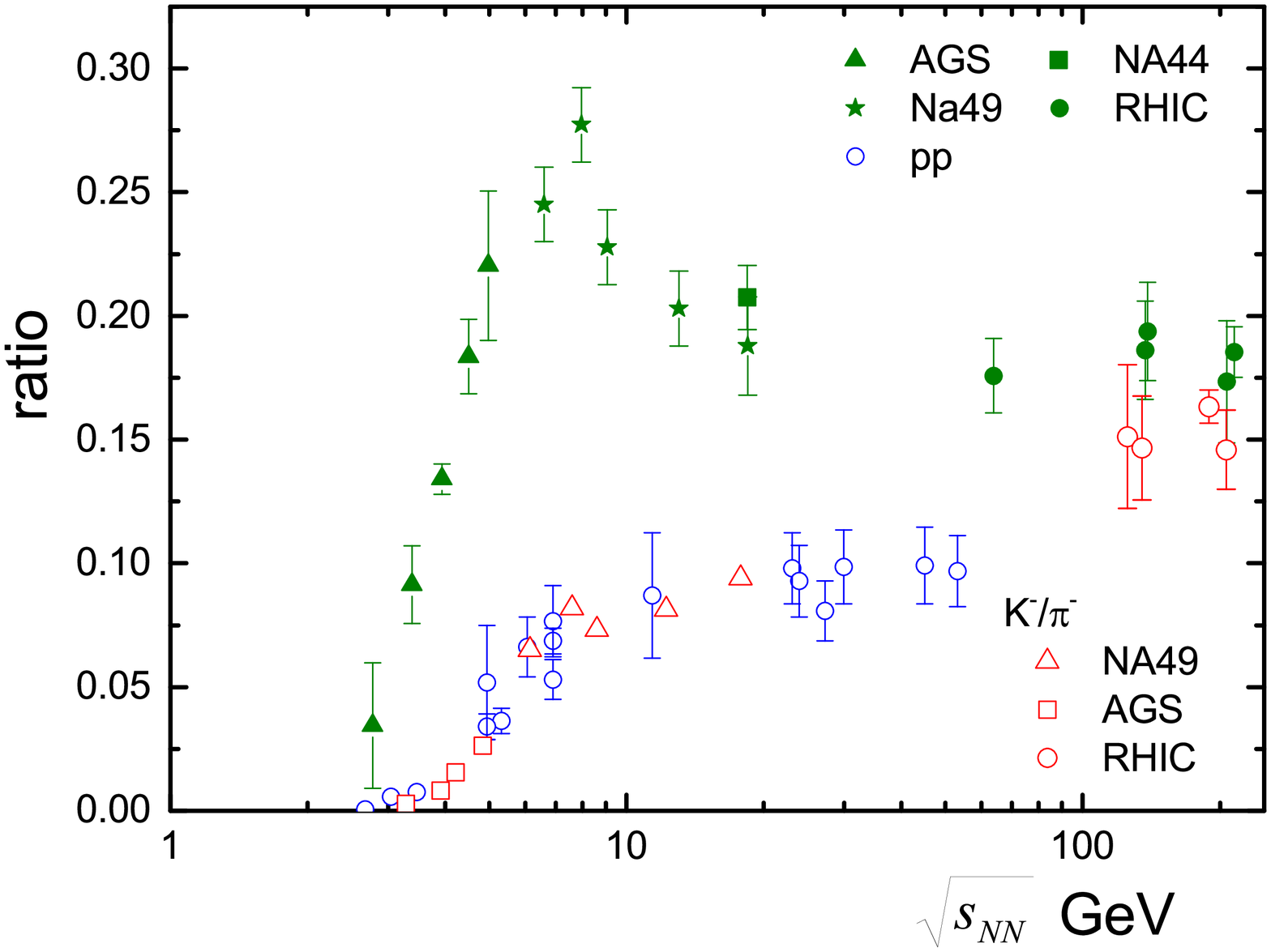}
}
\caption{The $K^+/\pi^+$ and $K^-/\pi^-$ ratio as the $\sqrt{s_{NN}}$  function for $Pb + Pb$ and $Au+Au$ central collisions \cite{NA49_exp,NA49_exp2,AGS_exp, RHIC_exp}. Blue circles are the  $K^+/\pi^+$ ratio for pp collisions.}
\label{Kpi_exp}
\end{figure}

The intriguing results were obtained at PbPb and AuAu collisions: a structure in the ratio of the  positive charged kaon  to  the positive charged pion named "horn" (see Fig. \ref{Kpi_exp})  was found. Firstly, the ''horn" was described by the NA49 Collaboration \cite{NA49_exp} and the work aroused a big interest. Investigations at energies from 7 to 200 AGeV were made at the  RHIC-BES (Beam energy scan) \cite{BES} and it was shown that data can be placed on the same curve \cite{NA49_exp,NA49_exp2,AGS_exp, RHIC_exp}. Such enhancement was also observed in the ratio of other positive charged strange particles to the positive charged non-strange pions. At that time, "horn" was not observed in the ratio of negative charged particles $K^-/\pi^-$ \cite{AdlerPLB595}. In the p+p collision the $K^+/\pi^+$ ratio shows smooth behaviour \cite{NA49_exp2,AdlerPLB595,pp_PRC69}.

The strangeness suppression in pp-collisions can  result from  insufficient  time and space of colliding and impossibility of reaching the statistical equilibrium of the strange flavour with the light quarks.

The exact theoretical reproduction of the "horn" in the $K/\pi$ ratio still does not exist. The microscopic transport model that includes only the hadron phase and does not include the quark-gluon phase can not reproduce experimental data \cite{artHSD, artHSD2, artUrQMD, artHGM, artRQMD}, and as a result, in the works \cite{NA49_exp2,onset_deconf} the authors suggested that such a peak in the ratio can be explained as the onset of deconfinement.

The SMES (the Statistical Model of Early Stages) considered a slow increase and the following jump 
in the ratio of the strange to non-strange particle production as a result of the deconfinement 
transition. According to the SMES, at low collision energies confined matter is produced, and the 
increase in the ratio is due the low T of the early stage and the large mass of strange particles. 
When the deconfinement transition occurs the strange quarks mass tends to its current mass ($m_s<T$) 
and the strangeness yield becomes independent of energy in the QGP (the "jump" from a high value of 
the ratio to its constant value) \cite{onset_deconf}.

The success in the desctiption of experimental data was achieved when the partial restoration of 
chiral symmetry  \cite{Bratkovskaya_chir} was added in the transport model. In the work the primary 
interaction was described through the mechanism of excitation and decay of color objects - strings. 
The function of the string fragmentation includes the dependence on the baryon density, thus 
modelling the mechanism of the partial chiral symmetry restoration 
\cite{Bratkovskaya_chir,CohenPRC}. The authors showed, that the partial chiral symmetry restoration 
is responsible for the quick increase in the  $K^+/\pi^+$ ratio at low energies and its decrease 
with the energy increasing (they explained this decreasing as a result of chiral condensate 
destruction).

The qualitative reproduction of the peak in the energy dependence of the kaon-to-pion ratio was 
obtained in the statistical model, that included hadron resonances and $\sigma$ - meson. This type 
of model implies the existence of the critical temperature for hadrons, which plays the role of the 
hadron phase transition  \cite{AndronicPLB}.

The SU(3) Nambu-Jona-Lasinio model with the Polyakov loop  seems to be most promising as an 
instrument for description of the chiral phase transition, the deconfinement properties and the 
existence of quarks and hadron states \cite{U_Fu08, CostaPRD79,EBlanquierJPG}. The chiral symmetry 
breaking in the model is explained through a mechanism of adding the chiral condensate to the 
current quark. The Polyakov loop extended model in addition to the chiral transition takes into 
account the deconfinement transition which is described by the Polyakov loop. The phase diagram of 
the PNJL model corresponds to its modern concept: at low temperature and high chemical potential the 
system suffers the 1st order phase transition. At high temperature and low chemical potential in 
system the chiral phase transition line is a crossover \cite{U_Fu08,EBlanquierJPG}. The disadvantage 
of the model is that the critical temperature of the crossover transition at low chemical potential 
in the PNJL model is higher than the one in the Lattice QCD $T_c =154(9)$\cite{Bazavov} and the 
temperature of critical end point (CEP) is lower than in others models \cite{Friesen2012, 
Friesen2015, Bazavov}.

We address this paper to the problem of kaon to pion ratio in the context of the SU(3) PNJL model. In Sec.II the formalism of the PNJL model and the behavior of mesons and quarks at zero chemical potential will be discussed. The PNJL model generalized to the finite chemical potential will be presented in Sec.III. In Sec.IV, the obtained  results are discussed and the conclusions are given.

\section{Model formalism}

We consider the Polyakov loop extended SU(3) Nambu-Jona-Lasinio model with scalar-pseudoscalar interaction and the t'Hooft interaction which breaks the U$_{A}(1)$ symmetry  \cite{CostaPRD79,EBlanquierJPG}. The global SU(3)$\otimes$SU(3) chiral symmetry of the Lagrangian is obviously broken by introduction of the nonzero current quarks mass $\hat{m}=\mbox{diag}(m_{u},m_{d},m_{s})$, and confinement/deconfinement properties (Z$_3$-symmetry) are described by the effective potential $\mathcal{U}(\Phi, \bar{\Phi}; T)$

\begin{eqnarray}
\mathcal{L\,} & = & \bar{q}\,(\,i\,{\gamma}^{\mu}\,D_{\mu}\,-\,\hat
{m} - \gamma_0\mu)\,q + \nonumber \\
&+&\frac{1}{2}\,g_{S}\,\,\sum_{a=0}^{8}\,[\,{(\,\bar{q}\,\lambda
^{a}\,q\,)}^{2}\,\,+\,\,{(\,\bar{q}\,i\,\gamma_{5}\,\lambda^{a}\,q\,)}%
^{2}\,]  \nonumber \\
& + & g_{D}\,\,\{\mbox{det}\,[\bar{q}\,(\,1\,+\,\gamma_{5}%
\,)\,q\,]+\mbox{det}\,[\bar{q}\,(\,1\,-\,\gamma_{5}\,)\,q\,]\,\} \nonumber\\
& - & \mathcal{U}(\Phi, \bar{\Phi}; T),
\label{lagr}%
\end{eqnarray}
where $q=(u,d,s)$ is the quark field with three flavours, $N_{f}=3$, and three colors, $N_{c}=3$, $\lambda^{a}$  are the Gell-Mann matrices,  $a =\displaystyle 0,1,\ldots,8$, ${\lambda^{0}=\sqrt{\frac{2}{3}}\,\mathbf{I}}$  and D$_\mu = \partial^\mu -i A^\mu$, where  $A^\mu$   is the gauge field with $A^0= -  iA_4$ and $A^\mu(x) = g_SA^\mu_a\frac{\lambda_a}{2}$  absorbs the strong interaction coupling.

The effective potential ${U}(\Phi, \bar{\Phi}; T$) depends on  temperature $T$, the Polyakov loop field $\Phi$ and its complex conjugated $\bar{\Phi}$, which can be obtained through the expectation value of the Polyakov line  \cite{PisarskiPRD62,RattiPRD73}:
\begin{equation}
\Phi(\overrightarrow{x}) =  \frac{1}{N_c}\rm{Tr}_c\langle\langle L(\overrightarrow{x})\rangle\rangle,
\end{equation}
where
\begin{equation}
L(\overrightarrow{x}) = \mathcal{P}\rm{exp}\left[i \int_0^\beta d\tau A_4(\overrightarrow{x},\tau)\right].
\end{equation}
The Polyakov loop field is the order parameter for $Z_3$-symmetry restoration, which is restored as   $\Phi \rightarrow 0$ (confinement) and broken as $\Phi\rightarrow 1$  (deconfinement) \cite{PolyakovPLB72}. The effective potential $U(\Phi,\bar{\Phi}; T)$ has to reproduce Lattice QCD data in the gauge sector \cite{Lat_U_boyd} and must satisfy the $Z_3$ center symmetry. Based on these suggestions one can choose any form of the potential \cite{RattiPRD73,U_Ratti_log,U_Fu08}. In this work the following general polynomial form is used \cite{RattiPRD73}:
\begin{eqnarray}\label{effpot}
&&\frac{\mathcal{U}\left(\Phi,\bar\Phi;T\right)}{T^4}= \nonumber\\
&&=-\frac{b_2\left(T\right)}{2}\bar\Phi \Phi-
\frac{b_3}{6}\left(\Phi^3+ {\bar\Phi}^3\right)+
\frac{b_4}{4}\left(\bar\Phi \Phi\right)^2,\ \nonumber \\ \label{Ueff}
&&b_2\left(T\right) = a_0+a_1\left(\frac{T_0}{T}\right)+a_2\left(\frac{T_0}{T}
\right)^2+a_3\left(\frac{T_0}{T}\right)^3~.
\end{eqnarray}
For the effective potential the following parameters were chosen: $T_0 = 0.19$ GeV, $a_0 =6.75$, $a_1 = -1.95$, $a_2 = 2.625$, $a_3 =-7.44$, $b_3 =0.75$, $b_4 =7.5$  \cite{RattiPRD73}.
The grand potential density for the PNJL model in the mean-field approximation can be obtained from the Lagrangian density(\ref{lagr}) \cite{KlevanskyRevMod64}:
\begin{eqnarray}
\Omega &=& U(\Phi, \bar{\Phi}; T) + g_S\sum_{i = u,d,s}\langle\bar{q}_iq_i\rangle ^2 + \nonumber\\
&+&4g_D\langle \bar{q}_u q_u\rangle\langle\bar{q}_d q_d\rangle\langle\bar{q}_s q_s\rangle  - 2N_c\sum_{i = u,d,s}\int\frac{d^3 p}{(2\pi)^3} E_i - \nonumber\\
&- &2 T\sum_{i = u,d,s}\int\frac{d^3 p}{(2\pi)^3} (N^+_\Phi(E_i) + N^-_\Phi(E_i))
\end{eqnarray}
with the functions
\begin{eqnarray}
N^+_\Phi(E_i)& = & {\rm Tr_c}\left[\ln(1+L^\dagger e^{-\beta(E_i-\mu)})\right] \nonumber\\
& = & \left[ 1+3\left( \Phi +\bar{\Phi} e^{-\beta
E_i^+}\right) e^{-\beta E_i^+} + e^{-3\beta E_i^+}
\right], \label{funcNp} \nonumber \\
\\
N^-_\Phi(E_i) & =& {\rm Tr_c}\left[\ln(1+L e^{-\beta(E_p + \mu)})\right]  \nonumber\\
&=& \left[ 1+3\left( \bar{\Phi} + {\Phi} e^{-\beta
E_p^-}\right) e^{-\beta E_p^-} + e^{-3\beta E_p^-} \right]~ \nonumber,\\
\label{funcNm}
\end{eqnarray}
where $E_i^\pm = E_i \mp \mu_i$, $\beta = 1/T$,  $E_{\rm i} = \sqrt{{\bf p_i}^2 + m_i^2}$  is the energy of quarks and $\langle\bar{q_i}q_i\rangle$ is the quark condensate.

To obtain the value of the Polyakov loop field  $\Phi$, $\bar{\Phi}$, one needs to minimize the grand potential  over its parameters
\begin{equation}
\frac{\partial \Omega}{\partial \Phi} = 0, \ \ \ \frac{\partial \Omega}{\partial \bar{\Phi}} = 0.
\end{equation}
The gap equation for quarks depends on the quark condensates:
\begin{equation}
m_i = m_{0i} - 2 g_S\langle\bar{q_i}q_i\rangle - 2g_D\langle\bar{q_j}q_j\rangle\langle\bar{q_k}q_k\rangle,
\label{gap_eq}
\end{equation}
where $i, j, k = $u, d, s are chosen in cyclic order. The quark condensate is \cite{CostaPRD79,KlevanskyRevMod64}:
\begin{eqnarray}
\langle\bar{q_i}q_i\rangle &=& i \int\frac{dp}{(2\pi)^4}\rm{Tr}S(p_i)= \nonumber\\
&=& - 2 N_c \int\frac{d^3p}{(2\pi)^3}\frac{m_i}{E_i}(1 - f^+_\Phi(E_i) - f^-_\Phi(E_i))\nonumber\\
\end{eqnarray}
with the modified Fermi functions:
\begin{eqnarray}
&&f^+_\P(E_{\rm p}- \mu)  = \nonumber\\
&&\frac{\AP e^{-\beta(E_{\rm p} - \mu)} + 2\P e^{-2\beta (E_{\rm p} - \mu)} + e^{-3\beta (E_{\rm p} - \mu)}}{ 1 + 3(\AP  + \P e^{-\beta (E_{\rm p} - \mu)})e^{-\beta (E_{\rm p} - \mu)} + e^{-3\beta (E_{\rm p} - \mu)}},\nonumber\\
\\
&&f^-_\P(E_{\rm p}+ \mu)  =  \nonumber\\
&&\frac{\P e^{-\beta(E_{\rm p} + \mu)} + 2\AP e^{-2\beta (E_{\rm p} + \mu)} + e^{-3\beta (E_{\rm p} + \mu)}}{ 1 + 3(\P  + \AP e^{-\beta (E_{\rm p} + \mu)})e^{-\beta (E_{\rm p} + \mu)} + e^{-3\beta (E_{\rm p} + \mu)}}.\nonumber\\
\label{modfermi}
\end{eqnarray}

The meson masses in NJL-like models are defined by the Bethe-Salpeter equation at $\mathbf{P}=0$ \cite{RKHPRC53}
\begin{equation}
1-P_{ij}\Pi_{ij}^{P}(P_{0}=M,\mathbf{P}=\mathbf{0})=0~, \label{rdisp}
\end{equation}
where for non-diagonal pseudo-scalar mesons $\pi\,,K$:
\begin{eqnarray}
P_{\pi}&=&g_{S}+g_{D}\left\langle\bar{q}_{s}q_{s}\right\rangle, \label{Ppi} \\
P_{K} &=& g_{S}+g_{D}\left\langle\bar{q}_{u}q_{u}\right\rangle, \label{Pkaon}
\end{eqnarray}
and the polarization operator has the form
\begin{equation}
\Pi_{ij}^{P}(P_{0})=4\left(  (I_{1}^{i}+I_{1}^{j})-[P_{0}^{2}-(m_{i}%
-m_{j})^{2}]\,\,I_{2}^{ij}(P_{0})\right), \label{ppij}
\end{equation}
where integrals  $I_{1}^{i}$ è $I_{2}^{ij}(P_{0})$ are defined as:
\begin{equation}
I_{1}^{i}=iN_{c}\int\frac{d^{4}p}{(2\pi)^{4}}\,\frac{1}{p^{2}-m_{i}^{2}}, \label{i1}
\end{equation}%
\begin{equation}
I_{2}^{ij}(P_{0})   =iN_{c}\int\frac{d^{4}p}{(2\pi)^{4}}\,\frac{1}%
{(p^{2}-m_{i}^{2})((p+P_{0})^{2}-m_{j}^{2})} \label{i2},
\end{equation}
with the quark energy $E_{i,j}=\sqrt{\mathtt{p}^{2}+m_{i,j}^{2}}$.  When the meson mass exceeds the total value of its consistent $P_0> m_i + m_j$, the meson turns into the resonance state. In this case, the complex properties of the integrals have to be taken into account and the solution has to be defined in the form $\displaystyle{P_0 = M_M - \dfrac{1}{2}i \Gamma_M}$. Each equation (\ref{rdisp}) splits into two equations from which the meson mass $M_M$ and the meson width $\Gamma_M$ can be obtained \cite{BlaschkeHorn}.

\begin{figure}[h]
\centerline{
\includegraphics[width = 6.5 cm]{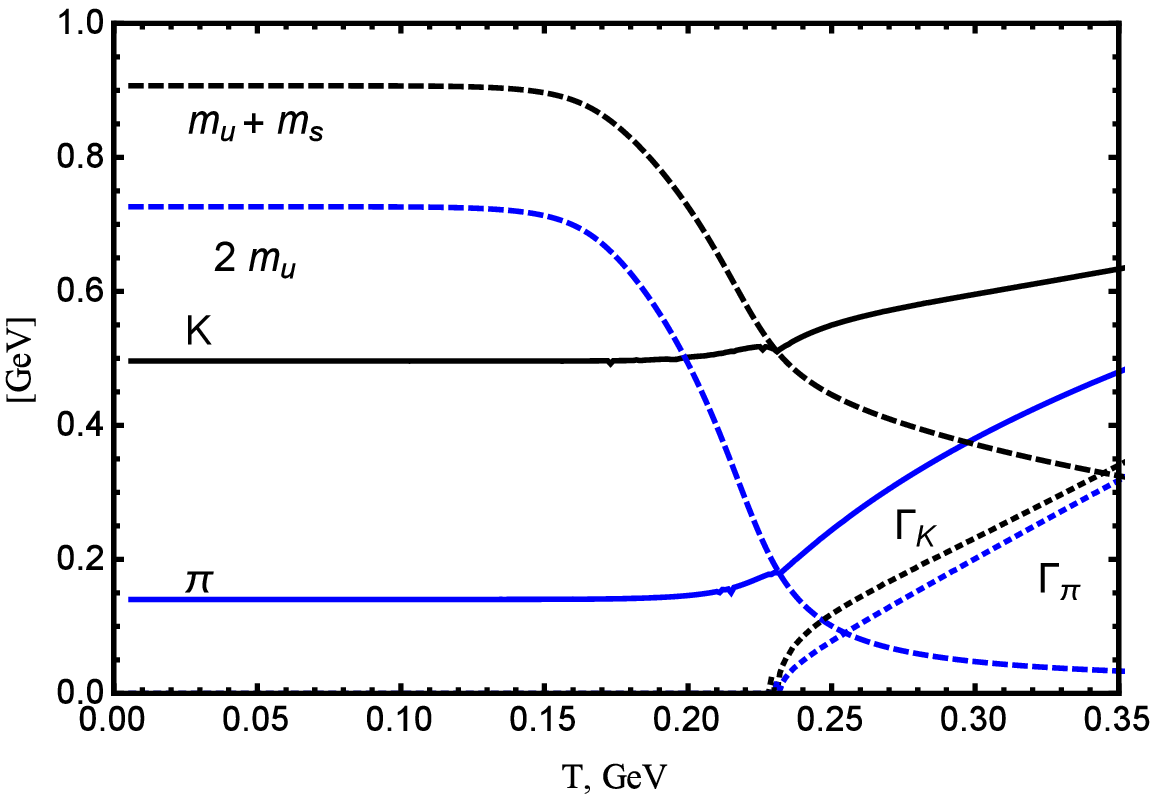}}
\centerline{
\includegraphics[width = 6.5 cm]{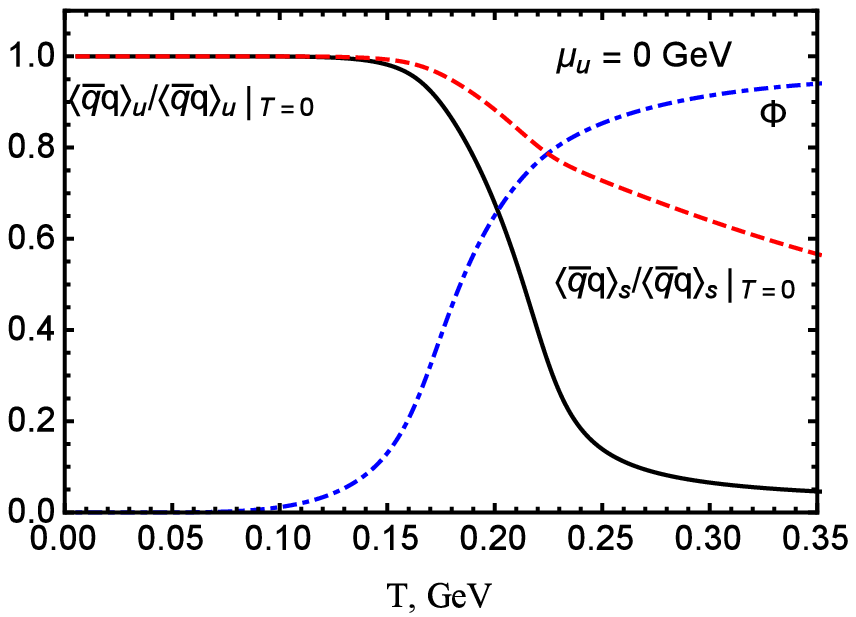}}
\caption{Top panel: The mass spectra at $\mu_u = \mu_d = \mu_s = 0$. The solid lines are denote the meson masses  Ê-meson (black) and pion (blue), dashed lines are denote $q\bar{q}$ the thresholds for the mesons, dotted lines are denote the meson width (color online). Bottom panel: the quark normalized condensates of light and strange quarks and the Polyakov loop field $\Phi$.}
\label{masses_T}
\end{figure}

The mass spectrum for the zero chemical potential is shown in Fig. \ref{masses_T}. The following set of parameters  was chosen for calculations: the current quark masses $m_{0 u} = m_{0 d} = 5.5$ MeV, $m_{0 s} = 0.131$ GeV,  the cut-off $\Lambda = 0.652$ GeV, couplings $g_D = 89.9$ GeV$^-2$ and $g_S = 4.3$ GeV$^{-5}$. As can be seen in Fig.\ref{masses_T}, at zero chemical potential charged multiplets are degenerated.  The $q\bar{q}$ threshold for the mesons is defined as $2m_u$ for pion and $m_u + m_s$ for kaon. The temperature at which the meson mass becomes equal to the value of the $q\bar{q}$ threshold is the Mott temperature $T^\pi_{Mott}$. After the Mott temperature the meson from the bound state turns into the resonance state and  can dissociate into its constituents.  As can be seen, the pion and kaon in the PNJL model decay at a near temperature ($T^\pi_{Mott} = 0.232$, $T^K_{Mott} = 0.23$ Gev).

The normalized quark condensate of light and strange quarks and the Polyakov loop field $\Phi$ are shown in the bottom panel Fig.\ref{masses_T}. The chiral condensate is the order parameter of the spontaneous chiral symmetry restoration in the model, and when the temperature exceeds a characteristic transition temperature, the condensate "melts" ($<q\bar{q}>\rightarrow 0$), the mass of quarks tends to their current values and the chiral symmetry is restored. At that time, the Polyakov loop field $\Phi\rightarrow 1$ and it is the signal of $Z_3$-symmetry breaking and deconfinement occurrence. As can be seen, the strange quark condensate is very high and the chiral symmetry restoration in the strange sector does not occur.

\section{Finite baryon density}

The calculations in SU(3) NJL-like models are complicated by the need to introduce the strange quark chemical potential. As a rule, when only thermodynamics is considered, the chemical potential of the strange quark is supposed equal to zero $\mu_s = 0$ GeV. We consider the following two cases :
\begin{itemize}
\item (Case I)  matter with equal chemical potentials  $\mu_u = \mu_d = \mu_s$
\item (Case II) with $\mu_u = \mu_d$, and $\mu_s = 0.55 \mu_u $
\end{itemize}
\begin{figure}[h]
\centerline{
\includegraphics[width = 7.5 cm]{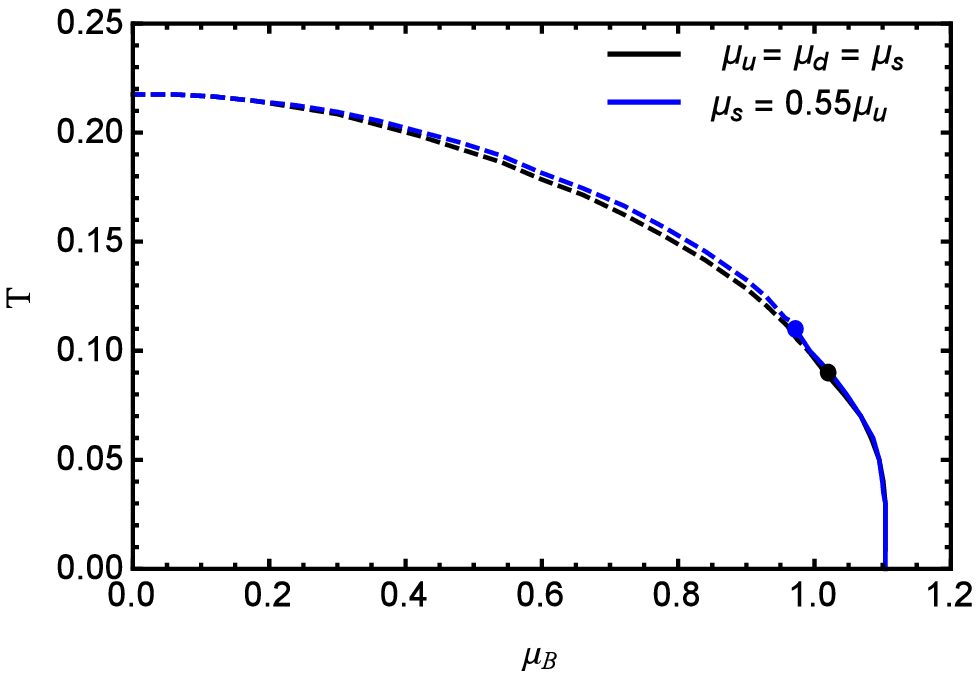}}
\centerline{
\includegraphics[width = 7.5 cm]{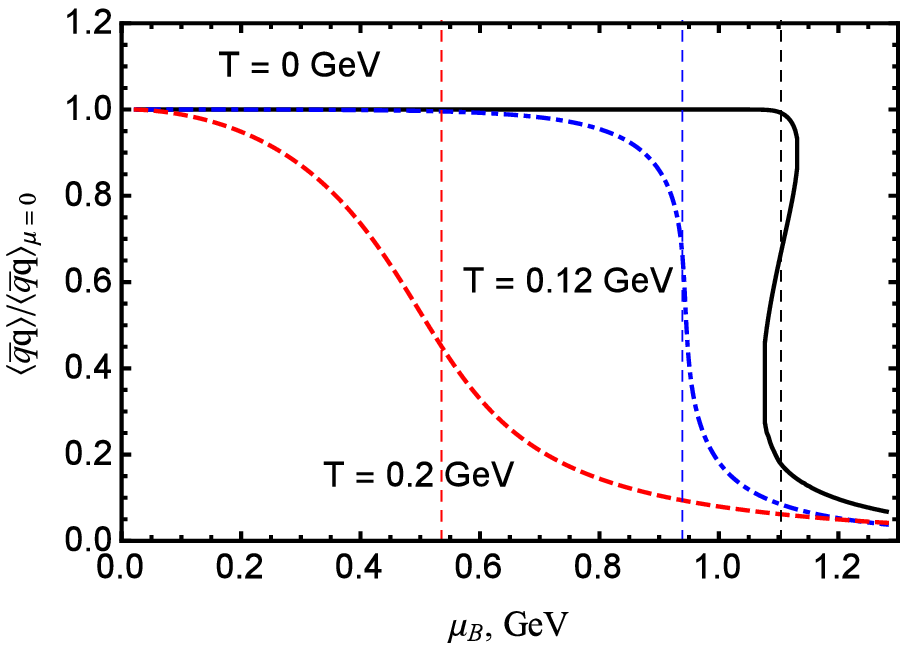}}
\caption{Top panel: the phase diagram in the T-$\mu_B$ plane. Bottom  panel: the normalized chiral condensate as a function of $\mu_B$ at different values of  $T = $0, 0.12, 0.2 GeV before, near and after CEP.}
\label{pd_muconst04}
\end{figure}

For both cases the phase diagrams are similar and are shown in Fig.\ref{pd_muconst04} (top panel). The phase diagram also has the structure similar to the SU(2) NJL case \cite{Friesen2012, Friesen2015}: at high temperature and low density (chemical potential) the phase transition is soft (crossover), the points of the crossover are defined as local maximum of $\dfrac{\partial <\bar{q}q>}{\partial T}|_{\mu_B = const}$. At low temperatures and high chemical potential crossover turns into the 1st order transition which can be defined as maximum of  $\dfrac{\partial^2 \Omega}{\partial \mu_u^2}|_{T = const}$ \cite{kunihiroPLB219}. The first order phase transition ends at the point called the critical endpoint (CEP)  ($\mu_{B, CEP} =$0.99, $T_{CEP}=$ 0.1) for the Case I and CEP ($\mu_{B, CEP} =$0.972, $T_{CEP}=$ 0.11) (for the case II), the position of CEP in both cases is  close to each other. As can be seen, the critical temperature of the crossover transition at $\mu_B = 0$ GeV is higher ($T_c = 0.218$) than it was predicted by the Lattice QCD $T_c =154(9)$  \cite{Bazavov}.

In Fig.\ref{pd_muconst04} (top panel), the chiral condensate as a function of the chemical potential is shown. As can be seen, at low temperature the gap equations (\ref{gap_eq}) have several solutions (or a break as a function $\mu_B$) and in the system there appears the 1st order transition; at high temperature (after CEP) the quark condensate changes softly  (crossover).

\begin{figure}[h]
\centerline{
\includegraphics[width = 7.1 cm]{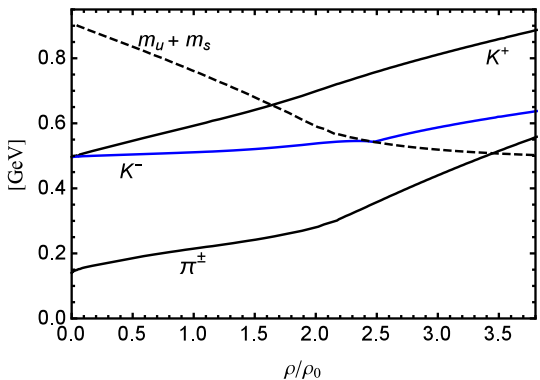}}
\caption{The spectra of meson masses as a function of the normalized baryon density at T = 0 GeV for Case (I).}
\label{Kaonmass_ro}
\end{figure}

At nonzero chemical potential and low T, the splitting of mass in charged multiplets is due excitation of the Dirac sea modified by the presence of the the medium (see Fig. \ref{Kaonmass_ro}). In dense baryon matter the concentration of light quarks is very high \cite{Stachel_ss}. Therefore, the creation of a $s\bar{s}$ pair dominates because of the Pauli principle: when Fermi energy for light quarks is higher than $s\bar{s}$-mass, the creation of the last one is energy-efficient.  The increase in the $K^+$  ($\bar{u}s$) mass, with respect to these of $K^-$ ($\bar{s}u$), is justified again by the Pauli blocking for s-quark (see for discussion \cite{Lutz,Ruivo_1996,CostaKalin_2003,CostaKalin_2004}).
Technically, to describe the mesons in dense matter, it is needed to  relate the chemical potential of quarks with Fermi momentum $\lambda_i$,  $\mu_i = \sqrt{\lambda_i^2 + m_i^2}$. The later affects  the limits of integration in of Eq. (\ref{i1}-\ref{i2}) . It is obviously that the pion for the chosen cases ($m_u = m_d$) is still degenerate.

To discuss the "horn" problem, we have to consider the ratio of the number of kaons to the number of pions. Within the PNJL model the number densities of mesons ($K^\pm/\pi^\pm = n_{K^\pm}/n_{\pi^{\pm}}$) can be calculated as  :
\begin{eqnarray}
n_{K^{\pm}} &=& \int_0^\infty p^2dp\frac{1}{e^{(\sqrt{p^2+m_{K^{\pm}}}\mp\mu_{K^{\pm}})}-1}, \\
n_{\pi^\pm} &=& \int_0^\infty p^2dp\frac{1}{e^{(\sqrt{p^2+m_{\pi^{\pm}}}\mp\mu_{\pi^{\pm}})}-1}.
\end{eqnarray}
The chemical potential for pions is a phenomenological parameter and in this work it was chosen as a constant $\mu_\pi = 0.135$ following the works \cite{pot_pi,BegunPRC90,naskret} but in \cite{BegunPRC90} it was supposed that $\mu_\pi$ can depend on T.  The chemical potential for kaons can be defined (see for example \cite{naskret,pot_K}) from $\mu_q = B_q\mu_B + S_q\mu_s + I_q\mu_q$, and in isospin   symmetry case ($I_q = 0$), the result is $\mu_K =\mu_u-\mu_s$.

\begin{figure}[h]
\centerline{
\includegraphics[width = 6.5 cm]{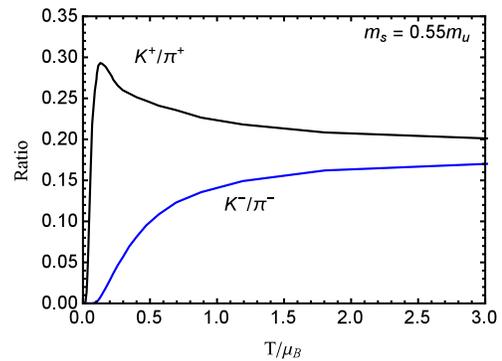}}
\centerline{
\includegraphics[width = 8.0 cm]{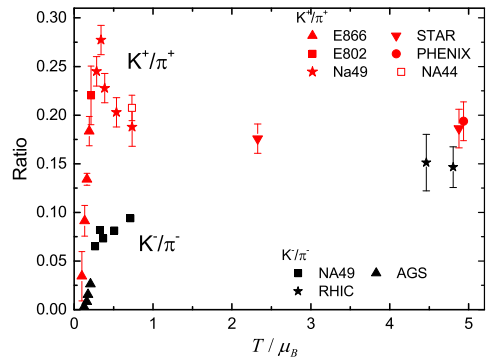}
}
\caption{Top panel: $K^+/\pi^+$, $K^-/\pi^-$ ratio as function of $T/\mu_B$ for different cases (I) and (II). Bottom panel: experimental data for $K^+/\pi^+$ ratio as function of rescaled variable $T/\mu_B$. }
\label{nKnpi_ratio}
\end{figure}

If  all experimental data are taken from various experiments, it is shown in the statistical model that for each experiment the temperature and the baryon chemical potential of freeze-out \cite{Cleymans_diagr} can be found using the parametrization suggested by J. Cleymans and al. It turned out possible to re-scale experimental data in variable $T/\mu_B$ (see Fig.\ref{nKnpi_ratio}, bottom panel) which is more suitable for theoretical calculation

\begin{eqnarray}
T(\mu_B) &=& a - b\mu^2_B- c\mu^4_B, \\
\mu_B(\sqrt{s})& = &\frac{d}{1 + e\sqrt{s}},
\label{param_exp}
\end{eqnarray}
where $a = 0.166 \pm 0.002$ GeV, $b = 0.139 \pm 0.016$ GeV$^{-1}$,
and $c = 0.053 \pm 0.021$ GeV$^{-3}$, $d = 1.308 \pm 0.028$ GeV, $e = 0.273 \pm 0.008$ GeV$^{-1}$.

It is evident that experimental data at higher energy correspond to high temperature and low density/chemical potential, and the data at low energy correspond to high density/chemical potential and low temperature.

The calculates (top panel) and re-scaled experimental (bottom panel) ratios $n_{K^\pm}/n_{\pi^{\pm}}$ are shown in Fig. \ref{nKnpi_ratio}(top panel) as a function of the scaling variable $T/\mu_B$, where values $T$ and $\mu_B$ were chosen along the chiral phase transition line,  which, generally speaking, does not coincide with the frieze-out curve.

It is clearly seen from the figure  that in the region of high temperature and low density (high values of $T/\mu_B$), the $K^+/\pi^+$ and $K^-/\pi^-$-ratios tend to the same value. These results are in agreement with experimental results. In the PNJL model at high temperature and low density the difference between the mass of charged kaons multiplets decreases, their masses become equal to each other (kaons are degenerate at T=0) and as can be seen, the difference in ratios can  also decrease.

At low values of $T/\mu_B$ (high chemical potential and low temperatures) the enhancement in the $K^+/\pi^+$ similar to experimental data is clearly seen. The absence of such structure in the $K^-/\pi^-$-ratio can be explained by that positive and negative charged kaons have different sensibility to the medium density. In time when the relative number of baryons decreases with increasing  energy, the number of negative charged kaons will not change, opposite to the positive charged kaons the number of which must be balanced by strange baryons and, therefore, will decrease \cite{Cleymans_Kminus}.

\begin{figure}[h]
\centerline{
\includegraphics[width = 6.5 cm]{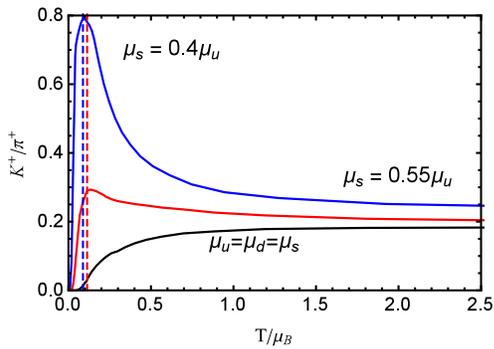}
}
\caption{The $K^+/\pi^+$ ratio as a function $T/\mu_B$ for different cases (I) and (II). }
\label{nKnpi_ratio_mu}
\end{figure}

\begin{figure}[h]
\centerline{
\includegraphics[width = 7 cm]{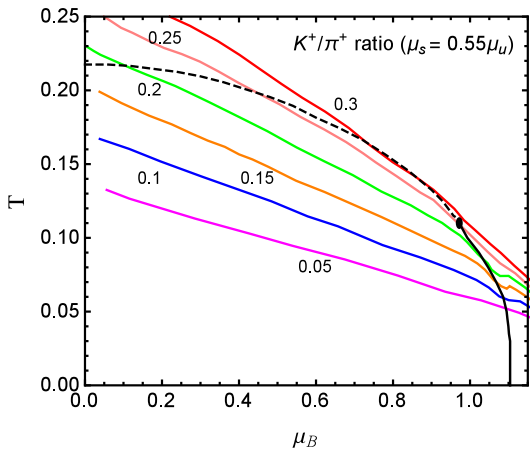}
}
\caption{The contour plot of the $K^+/\pi^+$-ratio on the phase diagram (black line) plane $T -\mu_B$. }
\label{contour}
\end{figure}

The quark matter properties in the frame of the PNJL model are formed by the choice of different assumptions for the environment: Case I and Case II. Generally speaking, both invented assumptions can not reproduce the properties of the medium in a real collision of heavy ions. Nevertheless, the choice of these two Cases can illustrate that the peak position is connected with the position of the critical end-point. In Fig.\ref{nKnpi_ratio_mu}, it can be seen how the enhancement in the $K^+/\pi^+$ ratio depends on the choice of the matter: in case (I), when  $\mu_K =0$ (and $\mu_S = 0$) there is no any enhancement in the ratio. For Case II it can be seen that the value of $\mu_s/\mu_u$ affects the position and the high of the peak in the kaon-to-pion ratio. As can be seen in Fig. \ref{nKnpi_ratio_mu}, the maxima are placed near the critical end points (vertical lines in Fig. \ref{nKnpi_ratio_mu}).  The high $K^+/\pi^+$ ratio also depends on the choice of the values $T$ and $\mu_B$ where it is calculated. The contour lines where the  $K^+/\pi^+$  ratio remains constant are shown in Fig.\ref{contour} on the phase diagram plane together with the phase diagram. As can be seen, the ratio reaches the maximal value near the critical point in the region of the 1st order transition and also slightly above the phase transition line. For comparison, similar lines were obtained in the statistical model \cite{Cleymans_contour} where the transition line is the line of freeze-out.

\section{Conclusion}

The strangeness enhancement in heavy ion collision has been suggested as the QGP signal long ago \cite{Rafelski1982,Koch1986}. The reason is that the probability of the process $gg\rightarrow s\bar{s}$ in QGP increases due the high density of gluons and due the chiral symmetry restoration in the strange sector (s-mass becomes lower). In dense matter the creation of the $s\bar{s}$ pair dominates because of the Pauli principle: high concentration of light quarks leads to that the Fermi energy for light quarks becomes higher, than the $s\bar{s}$-mass and the creation of the last one is energy-efficient. Therefore, if in the heavy ion collision the quark-gluon plasma is created, the enhancement of the strangeness can be expected in comparison soon with the p-p or p-N collision.

The dependence of the strangeness to non-strangeness ratio was discussed in Ref. \cite{Bratkovskaya_chir,Cleymans_contour}.  In the standard picture, freeze-out is interpreted as a competition between fireball expansion and interaction
of the constituents. Thus, it is natural to expect the system size dependence in freeze-out conditions, since constituent interactions decrease as one goes from nucleus-nucleus (A+A) to proton-nucleus (p+A) and proton-proton (p+p) collisions. The freeze-out hypersurface is usually extracted for three different freeze-out schemes that differ in the way strangeness is treated. The recent results~\cite{nucl-size} confirm expectations from the previous analysis of the system size dependence in the freeze-out scheme with mean hadron yields: while heavy ion collisions are dominated by constituent interactions, smaller collision systems like proton + nucleus and proton + proton collisions with lesser constituent interaction prefer a unified freeze-out scheme
with varying degree of strangeness equilibration.

The main idea of this work was to show that the cause of appearance of the "horn" at energies $\sqrt{s_{NN}} = 8-10$ GeV may be a qualitative change in the state of the environment where kaons and pions are created. In the work \cite{Bratkovskaya_chir} the quick increase in the $K^+/\pi^+$ ratio and its decrease  with further increasing energy was interpreted as a sequence of the chiral symmetry restoration effect and the deconfinement effect. In the PNJL model the picture is the following (Fig. \ref{pd_muconst04}): when $T$ and $\mu_B$ are chosen on the phase diagram line, the system is in the phase transition region and the chiral condensate is still not destroyed. The main difference between the choice of $T$ and $\mu_B$ along the line is whether we are in the crossover region or in the 1st order transition region (vertical lines in Fig.\ref{pd_muconst04}, bottom panel). In the region of the 1st order transition (low temperatures) the value of $\Phi \rightarrow 0$ and the matter is confined. In the region of the crossover deconfinement transition takes place.

In the PNJL model, the masses of positive and negative mesons are splitted at high densities.  This splitting can explain the difference in  the behavior of the $ K /\pi $ ratios for different charge signs in the high-density region and the fact that they tend to the same value at high temperatures and low densities, where kaons become degenerate. It was shown that in case (I) (when the chemical potential of the strange quark coincides with the chemical potential of the light quark and there is no strange chemical potential in the system), the enhancement in the $K^+/\pi^+$ ratio is absent. This can be a signal that the peak is sensitive to the properties of the matter. Therefore, as a future work, one can check the presence/absence of such a peak in different media (a medium with an equal baryon density, a medium with beta equilibrium, strange matter, or a medium with an equal number of protons of neutrons and hyperons, etc.). The secon interest result was that the region of maximum values in the ratio is localized near the critical end point. This hypothesis can also be verified by including a vector interaction in the PNJL model, which can allow one to move the critical point up to its removing from the phase diagram.

\section{Acknowledgements}

We are thankful to A. Khvorostukhin and E. Kolomeitsev  for useful advice. We also thank W. Cassing for the discussion. The work  A.F. was supported by the Russian Science Foundation under grant no. 17-12-01427.


\begin{thebibliography}{100}

\bibitem{NA49_exp}
S. V. Afanasiev {\it et al.} [NA49 Collabration],
Phys. Rev. C {\bf 66}, 054902 (2002).

\bibitem{BES}
STAR BES White paper, SN0598.

\bibitem{NA49_exp2}
C. Alt,{\it et al} (NA49 Collaboration),
Phys. Rev.  C {\bf 77}, 024903 (2008).

\bibitem{AGS_exp}
 J. L. Klay et al. (E895 Collaboration),
 Phys. Rev.  C  {\bf 68}, 054905 (2003).

\bibitem{RHIC_exp}
B. I. Abelev et al. (STAR Collaboration), Phys. Rev. {\bf C 81}, 024911. (2010); M. M. Aggarwal et al. (STAR Collaboration), ibid. {\bf 83}, 024901 (2011).


\bibitem{AdlerPLB595}
C. Adler et al. [STAR Collaboration],
Phys. Lett.  B {\bf 595} (2004) 143; arXiv: 0206008 .


\bibitem{pp_PRC69}
M. Kliemant, B. Lungwitz, and M. Ga\'zdzicki,
Phys. Rev. C {\bf 69}, 044903 (2004).

\bibitem{artHSD}
W. Ehehalt and W. Cassing,
Nucl. Phys. A {\bf 602}, 449 (1996);

\bibitem{artHSD2}
 W. Cassing and E. L. Bratkovskaya,
 Phys. Rep. {\bf 308}, 65 (1999).

\bibitem{artUrQMD}
H. Petersen, M. Bleicher, s. A. Bass, H. Stocker,
arXiv:0805.0567 [hep-ph].

\bibitem{artHGM}
J. Cleymans, E. Suhoen, G. M. Weber,
Z. Phys.  C {\bf 53}, 485 (1992).

\bibitem{artRQMD}
H. Sorge,
Phys. Rev. C {\bf  52}, 3291 (1995).

\bibitem{onset_deconf}
M. Gazdzicki, M.I. Gorenstein, Acta Phys. Pol. B {\bf  30}, 2705 (1999)
[arXiv:hep-ph/9803462]; {\bf  42} (2011) 307.

\bibitem{Bratkovskaya_chir}
A. Palmese, W. Cassing, F. Seifer, T. Steinert, P. Moreau and E.L. Bratkovskaya,
Phys. Rev. C {\bf  94}, 044912 (2016).

\bibitem{CohenPRC}
T. D. Cohen, R. J. Furnstahl and D. K. Griegel,
Phys. Rev. C {\bf  45}, 1881 (1992).

\bibitem{AndronicPLB}
 A. Andronic, P. Braun-Munzinger, J. Stachel,
 Phys. Let. B {\bf 673}, 142-145 (2009).

\bibitem{U_Fu08}
K. Fukushima,
Phys. Rev. D {\bf  77}, 114028 (2008).
\bibitem{CostaPRD79}	
P. Costa, M. C. Ruivo, C. A. de Sousa, H. Hansen and W. M. Alberico
Phys. Rev. D {\bf 79}, 116003 (2009).
	
\bibitem{EBlanquierJPG}	
E. Blanquier
J. Phys. G: Nucl. Part. Phys. {\bf  38}, 105003 (2011).


\bibitem{Bazavov}
A. Bazavov et al. (HotQCD Collaboration),
Phys. Rev. D {\bf 85}, 054503 (2012).


\bibitem{Friesen2012}
A. V. Friesen, Yu. L. Kalinovsky, V. D. Toneev,
Int. J. Mod. Phys. A 27, {\bf  1250013}, (2012).

\bibitem{Friesen2015}
A. V. Friesen, Yu. L. Kalinovsky, V. D. Toneev,
Int. J. Mod. Phys. A {\bf 30}, 1550089 (2015).


\bibitem{PisarskiPRD62}
R. D. Pisarski,
Phys. Rev. D {\bf 62}, 111501 (2000); arXiv: hep-ph/0203271.

\bibitem{RattiPRD73}
C. Ratti, M. A. Thaler, and W. Weise,
Phys. Rev. D {\bf 73}, 014019 (2006); arXiv:nucl-th/0604025

\bibitem{PolyakovPLB72}
A. M. Polyakov,
Phys. Lett. B {\bf 72}, 477 (1978).

\bibitem{Lat_U_boyd}
G. Boyd  {\it et. al.},
Nucl. Phys. B {\bf 469}, 419 (1996).

\bibitem{U_Ratti_log}
S. R\"ossner, C. Ratti, W. Weise,
Phys. Rev. D {\bf 75}, 034007 (2007).

\bibitem{KlevanskyRevMod64}
S. P. Klevansky, Rev. Mod. Phys. {\bf 64}, 649 (1992).

\bibitem {RKHPRC53}
	P. Rehberg, S. P. Klevansky and J. H\"ufner,
	Phys. Rev. C \textbf{53}, 410 (1996).
	
\bibitem{BlaschkeHorn}
A. Dubinin, A. Radzhabov, D. Blaschke, A. Wergieluk,
Phys. Rev. D {\bf 96}, 094008 (2017).


\bibitem {kunihiroPLB219}
T. Kunihiro,
Phys. Lett. B \textbf{219}, 363 (1989).


\bibitem{Stachel_ss}
J. Stachel, G. R. Young,
Rev. Nucl. Part. Sci.  {\bf 42} 537 (1992) .

\bibitem{Lutz}
M. Lutz, A. Steiner, W. Weise,
Nucl. Phys. A, {\bf 574} 755 (1994).

\bibitem{Ruivo_1996}
M.C. Ruivo, C.A. de Sousa,
Phys. Let. B {\bf 385}, 39 (1996).

\bibitem{CostaKalin_2003}
P. Costa, M. C. Ruivo, Yu. L. Kalinovsky,
Phys. Lett. {\bf B560}, 171 (2003) .


\bibitem{CostaKalin_2004}
P. Costa, M. C. Ruivo, Yu. L. Kalinovsky, C. A. de Sousa,
Phys. Rev. C {\bf 70} (2004) 025204.

\bibitem{pot_pi}
M. Kataja, P.V. Ruuskanen
Phys. Let. B {\bf 243}, 181 (1990).

\bibitem{BegunPRC90}
V. Begun, W. Florkowski, M. Rybczynski,
Phys. Rev.  C  {\bf 90}, 014906 (2014).


\bibitem{naskret}
M. Naskret, D. Blaschke, A. Dubinin,
Phys. Elem. Part. Atom. Nucl. {\bf 46}, 1445 (2015).

\bibitem{pot_K}
A. Lavagno and D. Pigato,
EPJ Web of Conferences {\bf 37}, 09022 (2012)

\bibitem{Cleymans_diagr}
J. Cleymans, H. Oeschler, K. Redlich, S. Wheaton,
Phys. Rev. C  {\bf 73}, 034905 (2006)

\bibitem{Cleymans_Kminus}
J. Cleymans, H. Oeschler, K. Redlich, S. Wheaton,
arXiv:hep-ph/0504065;

\bibitem{Cleymans_contour}
H. Oeschler, J. Cleymans, K. Redlich, S. Wheaton,
PoS CPOD2009:032,2009; arXiv:1603.09553




\bibitem{Rafelski1982}
Rafelski J., Muller B.,
Phys. Rev. Lett. {\bf 48}, 1066-1069 (1982).

\bibitem{Koch1986}
Koch P., Muller B., Rafelski J.
Phys. Rep. {\bf 142},  167 (1986).

\bibitem{nucl-size}
A. K. Dash,  R. Singh, S. Chatterjee, Ch.n Jena, B. Mohanty,
arXiv:1807.06829.


\end{thebibliography}
\end{document}